\documentclass[aps,amsfonts,nofootinbib,article,twocolumn,showpacs]{revtex4}
\usepackage{amsmath}
\usepackage{amsfonts}
\usepackage{amssymb}

\def\openone{\leavevmode\hbox{\small1\kern-3.8pt\normalsize1}}
\def\N{\leavevmode\hbox{ Z \kern-8 pt\normalsize{Z}}}
\def\openone{\leavevmode\hbox{\small1\kern-3.8pt\normalsize1}}
\def\openJ{\leavevmode\hbox{J \kern-9.5pt\normalsize J}}
\def\openS{\leavevmode\hbox{ S \kern-9.3pt\normalsize S}}
\newcommand{\bb}{\begin{equation}}
\newcommand{\ee}{\end{equation}}
\newcommand{\eqb}{\begin{eqnarray}}
\newcommand{\eqf}{\end{eqnarray}}

\usepackage{color}

\begin{document}

\title{Can gravitation accelerate neutrinos?}

\author{Sergio A. Hojman}
\email{sergio.hojman@uai.cl} \affiliation{Departamento de Ciencias,
Facultad de Artes Liberales, Facultad de Ingenier\'{\i}a y Ciencias,
Universidad Adolfo Ib\'a\~nez, Santiago, Chile,\\ and Departamento
de F\'{\i}sica, Facultad de Ciencias, Universidad de Chile,
Santiago, Chile,\\ and Centro de Recursos Educativos Avanzados,
CREA, Santiago, Chile.}

\author{Felipe A. Asenjo}
\email{faz@physics.utexas.edu}
\affiliation{Institute for Fusion Studies, The University of Texas
at Austin, Austin, Texas 78712, USA.}

\begin{abstract}
The Lagrangian equations of motion for massive spinning test particles (tops) moving on a gravitational background using General Relativity are presented. The paths followed by tops are nongeodesic. An exact solution for the motion of tops on a Schwarzschild background which allows for superluminal propagation of tops is studied. It is shown that the solution becomes relevant for particles with small masses, such as neutrinos. This general result is used to calculate the necessary condition to produce superluminal motion in part of the trajectory of a small mass particle in a weak gravitational field. The condition for superluminal motion establishes a relation between the mass, energy and total angular momentum of the particle.
\end{abstract}

\pacs{04.20.Cv, 04.20.Jb, 04.90.+e, 14.60.St, 14.60.Lm}

\maketitle
\section{Introduction}

The well known September 2011 OPERA neutrino experiment \cite{OPERA} has produced
a myriad of articles trying to explain or understand, prove or
disprove the  experimental results for superluminal neutrino motion. One of the most important contributions to the understanding of the problem is the Cohen-Glashow work \cite{cg}.

Most of
articles are based on models which, by construction, ignore or neglect
general relativistic effects. Furthermore, even though the general
relativistic aspects of the dynamics are not considered, the
arguments presented in these papers seem to tacitly assume that
neutrinos follow geodesics in the presence of gravitational fields.

It has been known for quite some time \cite{mat,pap1,hojman1,hojman2} that spinning massive test
particles (tops) follow non geodesic paths when moving on gravitational fields.

Even though a recent communication seems to refute the
fact that the original OPERA experimental results do indeed describe
superluminal propagation of neutrinos, we believe that is
interesting to show that a consistent general relativistic
model allows for the possibility of superluminal propagation of
tops moving on gravitational backgrounds \cite{hojman1,hojman2,hojman2a}.

In this article we discuss an exact solution for the equations
describing the motion of a top on a Schwarzschild background. A
Lagrangian derivation for these non geodesic equations of motion
obeyed by tops moving on a gravitational background was first
obtained by Hojman \cite{hojman1,hojman2} (using a formalism by
Hanson and Regge \cite{hr} developed for flat spacetime) which it is
outlined in the next section. We show that massive tops may reach
superluminal velocities provided that their particle mass $m$, energy $E$
and total angular momentum magnitude $j$ of the orbit, satisfy a
relation (which is explicitly derived in Section IV).

Non geodesic equations of motion for tops were first derived by
Mathisson \cite{mat} and Papapetrou \cite{pap1} as limiting cases
of rotating fluids moving in gravitational fields. On the other
hand, massless spinning particles do follow null geodesics as showed
by Mashhoon \cite{mas} who used the Mathisson--Papapetrou formalism
for his derivation.

Loosely speaking, in the presence of gravitational fields, massive
spinning particles (such as neutrinos) follow non geodesic paths
\cite{mat,pap1,hojman1,hojman2} while photons move in null geodesics
(in spite of their spinning nature) because they are massless
\cite{mas}.

Even though the Equivalence Principle is sometimes interpreted as
stating that test particles in a gravitational field should follow
geodesics, this interpretation is, strictly speaking, valid only for
spinless point test particles. Extended particles are, in general,
subject to tidal forces and follow, therefore, non geodesic paths.

In the Lagrangian formulation of the motion of tops, the velocity
$u^\mu$ and the canonical momentum $P^\mu$ vectors are, in general,
not parallel. For the motion of tops in the presence of
electromagnetic and/or gravitational fields, the (square of the)
mass $m^2(\equiv P^\mu P_\mu> 0)$ is conserved implying that the
momentum vector remains timelike along the motion. Nevertheless, the
velocity vector may become spacelike \cite{hojman1,vz,hr,hojman3}.

The proper treatment of the
(subtle but crucial) lack of parallelism between velocity and
momentum is best achieved with a Lagrangian formulation of the
motion of tops, because otherwise the canonical momentum cannot be
appropriately defined. Furthermore, the Mathisson--Papapetrou
formulation gives rise to third order equations of motion, while the
Lagrangian approach gives rise to second order ones \cite{hojman1,
rasband}.

It turns out that for suitable choices of initial conditions this non geodesic motion can accelerate tops (and neutrinos)
beyond the speed of light.

Theoretical results involving superluminal propagation of massive
spinning particles and fields in interaction with electromagnetic or
gravitational fields have been reported previously in the literature
by Velo and Zwanziger \cite{vz}, Hanson and Regge \cite{hr}, Hojman
\cite{hojman1} and Hojman and Regge \cite{hojman3}, for instance,
while experiments reporting hints of superluminal neutrino
propagation can be found in Refs.~\cite{capa,giani}, among others,
and other results reported in the MIT webpage on Superluminal
Neutrinos (The Net Advance of Physics,
http://web.mit.edu/redingtn/www/netadv).

\section{Motion of a relativistic top on a gravitational field}

Consider a relativistic (spherical) top. Denote its position by
a four vector $x^\mu$ while its orientation is defined by an
orthonormal tetrad ${e_{(\alpha)}}^{\mu}$. A gravitational field is
described as usual in terms of the metric field $g_{\mu \nu}$
\cite{hojman1,hojman2}. The tetrad vectors satisfy
\begin{equation}
 g_{\mu \nu} \ {e_{(\alpha)}}^{\mu} \ {e_{{(\beta)}}}^{\nu} \ \equiv\ \eta_{(\alpha \beta)}\, , \label{eta1}
\end{equation}
with $\eta_{(\alpha \beta)} (= \eta^{(\alpha \beta)})$ given by
\begin{equation}
\eta_{(\alpha \beta)} \ \equiv \ \ \mbox{diag}\ (+1, -1, -1, -1)\, .
\label{eta2}
\end{equation}
and have, therefore, six independent components (consistent with the
number of parameters of the Lorentz group).

The velocity vector $u^\mu$ is defined in terms of an arbitrary
parameter $\lambda$ by
\begin{equation}
u^\mu\equiv \frac{d x^\mu}{d \lambda}\, . \label{vel}
\end{equation}

Besides, the antisymmetric angular velocity tensor $\sigma^{\mu \nu}$ is
\begin{equation}
\sigma^{\mu \nu}\ \equiv \eta^{(\alpha \beta)}
{e_{(\alpha)}}^{\mu}\frac{D{e_{{(\beta)}}}}{D \lambda} ^{\nu}\ = \ -
\ \sigma^{\nu \mu}\, ,\label{sigma}
\end{equation}
where the (covariant) derivative $D{e_{{(\beta)}}}^{\nu}/D\lambda $ is
defined in terms of the Christoffel symbols ${\Gamma^{\nu}}_{\rho
\tau}$ , as usual, by
\begin{equation}
\frac{D{e_{{(\beta)}}}}{D \lambda} \ \equiv \
\frac{d{e_{{(\beta)}}}}{d \lambda}\ + {\Gamma^{\nu}}_{\rho \tau} \
{e_{{(\beta)}}}^{\rho}\ u^\tau\, . \label{covder}
\end{equation}

Note that general covariance may be achieved unambiguously at the
level of the Lagrangian formulation \cite{hojman1} because only
first derivatives of the dynamical variables are used in the
construction of the Lagrangian. If implemented using the equations
of motion, terms proportional to the Riemann tensor may be missed,
because (flat spacetime) second partial derivatives commute while
(curved spacetime) second order covariant derivatives do not commute
and their commutator is proportional to the Riemann tensor. If no
Lagrangian theory for a system of special relativistic equations of
motion is known, the introduction of gravitational interactions
cannot be unambiguously implemented.

The Lagrangian $L = L (a_1, a_2, a_3, a_4)$ is constructed as an
arbitrary function of four invariants $a_1, a_2, a_3, a_4$ such that
the action $S=\int L\, d\lambda$, be $\lambda$--reparametrization
invariant
\begin{equation}
L (a_1, a_2, a_3, a_4) = (a_1)^{1/2}\mathcal{L} \left(a_2/a_1,
a_3/(a_1)^2, a_4/(a_1)^2\right)\, , \label{lag1}
\end{equation}
(the speed of light $c$ is set equal to $1$, at this time),
$\mathcal{L}$ is an arbitrary function of three variables and $
a_1\equiv u^\mu u_\mu,\ a_2\equiv \sigma^{\mu \nu} \sigma_{\mu \nu}=
- \mbox{tr}({\sigma}^2),\ a_3\equiv u_\alpha \sigma^{\alpha \beta}
\sigma_{\beta \gamma} u^\gamma,\ a_4\equiv \mbox{det}({\sigma})$.
Note that this expression seems to require $a_1$ to be positive. To
be precise, the fact that $a_1$ be positive is not crucial in the
formulation, as it will be proved later on [see Eq.~\eqref{lagsp0}].
One could rewrite the expression for the Lagrangian as
\begin{equation}
L (a_1, a_2, a_3, a_4) = (a_2)^{1/2}\mathcal{L}_1 (a_1/a_2,
a_3/(a_2)^2, a_4/(a_2)^2)\, , \label{lag2}
\end{equation}
for instance.

The conjugated momentum vector $P_\mu$ and antisymmetric spin tensor
$S_{\mu \nu}$ are defined by
\begin{equation}
P_\mu \equiv \frac{\partial L}{\partial u^\mu}\, ,\label{pmu}
\end{equation}
\begin{equation}
S_{\mu \nu} \equiv \frac{\partial L}{\partial \sigma^{\mu \nu}} = -
S_{\nu \mu}\, .\label{sigmamunu}
\end{equation}

The equations of motion are obtained by considering the variation of
the action $S$ with respect to (ten) independent variations $\delta
x^\mu$ and (the covariant generalization of) $\delta \theta^ {\mu
\nu}$ defined by
\begin{equation}
\delta \theta^{\mu \nu} \equiv \eta^{(\alpha \beta)}
{e_{(\alpha)}}^{\mu}{\delta {e_{{(\beta)}}}}^\nu\ = - \delta
\theta^{\nu \mu}\, .\label{thetamunu}
\end{equation}

The (non geodesic) equations of motion turn out to be \cite{hojman1,
hojman2}
\begin{equation}
 \frac{D P^\mu}{D\lambda}=-\frac{1}{2}{R^\mu}_{\nu\alpha\beta}u^\nu S^{\alpha\beta}\, ,
\label{momentummotion}
\end{equation}
and
\begin{equation}
\frac{D S^{\mu \nu}}{D\lambda}=S^{\mu
\lambda}{\sigma_\lambda}^\nu-\sigma^{\mu
\lambda}{S_\lambda}^\nu=P^\mu u^\nu-u^\mu P^\nu\, . \label{spinmotion}
\end{equation}

These results hold for arbitrary $\cal L$. The dynamical variables
$P^\mu$ and $S^{\mu \nu}$ may be interpreted as the ten generators
of the Poincar\'e group. In order to restrict the spin tensor to
generate rotations only, the Tulczyjew constraint
\cite{tulc}
\begin{equation}
S^{\mu \nu} P_\nu=0\, ,\label{constraint}
\end{equation}
is usually imposed \cite{hojman1, hr}. It turns out that both the
top mass $m$ and its spin $J$ are conserved quantities (see Appendix
A)
\begin{equation}
m^2\equiv P^\mu P_\mu\, , \label{mass}
\end{equation}
\begin{equation}
J^2\equiv \frac{1}{2} S^{\mu \nu}S_{\mu \nu}\, . \label{spin}
\end{equation}

Furthermore, if $\xi^\mu$ is a Killing vector, then
\begin{equation}
C_\xi\equiv P^\mu \xi_\mu-\frac{1}{2}S^{\mu \nu}\xi_{\mu;\nu}\, ,
\label{cxi}
\end{equation}
is a constant of motion \cite{hojman1,hojman2,hojman3}.

The following gauge choices and ``invariant relations''
(defined in Appendix B)
\begin{equation}
x^0 =\lambda=t\, ,\quad\qquad {e_{(0)}}^{\mu}=P^\mu/m\, ,
\label{gauge2}
\end{equation}
may be implemented \cite{hr,hojman1,hojman2} to fix the arbitrary
parameter $\lambda$ and to restrict the (Lorentz transformations)
six degrees of freedom of the tetrad to three dimensional rotations
(for details, please see Appendix  B). The previous
choices satisfy condition \eqref{mass}
\begin{equation}
{e_{(0)}}^\mu {e_{(0)}}_\mu=\frac{P^\mu P_\mu}{m^2}=\eta_{(00)}=1\,
,
\end{equation}
and are consistent with constraint \eqref{constraint}
\begin{eqnarray}\label{consist}
  \frac{D\left(S^{\mu\nu}P_\nu\right)}{D\lambda}&=&0\nonumber\\
  &=&\left(S^{\mu\lambda}{\sigma_\lambda}^\nu-\sigma^{\mu\lambda}{S_\lambda}^\nu\right)P_\nu+m S^{\mu\nu}\frac{D {e_{(0)}}_\nu}{D\lambda}\nonumber\\
  &=&m S^{\mu\lambda}{\sigma_\lambda}^\nu {e_{(0)}}_\nu-m S^{\mu\nu}\sigma_{\nu\lambda} {e_{(0)}}^\lambda=0\, ,
\end{eqnarray}
where we have used the fact that
\begin{equation}\label{tetradev}
\frac{D{e_{{(\alpha)}}}^{\mu}}{D \lambda} = -
{\sigma^{\mu}}_\lambda{e_{(\alpha)}}^\lambda\, ,
\end{equation}
for any of the tetrad vectors, according to the Eqs.~\eqref{sigma},
\eqref{eta1} and \eqref{eta2}.

The consistency of the constraint \eqref{constraint} with the
equations of motion \eqref{momentummotion} and \eqref{spinmotion} is
guaranteed by making use of the arbitrariness of Lagrangian
$\mathcal L$ (or $\mathcal L_1$) in \eqref{lag1} (or \eqref{lag2})
by appropriately constructing it. As a matter of fact, the
Lagrangian \cite{hr}
\begin{eqnarray}\label{lagsp0}
  L&=&\left(\frac{A a_1-B a_2}{2}\right.\nonumber\\
  &+&\left.\frac{1}{2}\sqrt{(A a_1-B a_2)^2-8B(A a_3-2B a_4)}\right)^{1/2}\, ,
\end{eqnarray}
gives rise to the equations of motion \eqref{momentummotion} and
\eqref{spinmotion} and the Tulczyjew constraint \eqref{constraint}
plus a Regge trajectory defined by $B m^2- A J^2/2= AB$. Besides, it can
 be proved this Lagrangian is well defined for $a_1\leq
0$. Therefore, one may consider Lagrangian \eqref{lagsp0} as the
starting point of this theory.

In order to prove that \eqref{lagsp0} is a well defined Lagrangian for
any value of $a_1$, define
\begin{equation}
D \equiv \frac{1}{2}\sqrt{(A a_1-B a_2)^2+16 B^2 a_4-8 A B a_3}\, ,
\label{root2}
\end{equation}
such that the Lagrangian \eqref{lagsp0} now becomes
\begin{equation}\label{lagsp1}
L=\left(\frac{A a_1-B a_2}{2} + D\right)^{1/2}\, .
\end{equation}
In what follows we show that $D \geq (A a_1-B a_2)/{2}$ which
ensures that $L$ is real for any value of $a_1$ under very general
assumptions.

First, we can see that $a_4$ is always non negative because the determinant of an
antisymmetric matrix in an even dimensional space is a perfect
square. The sign of $a_3$ can be found as follows. We have that
\begin{equation}
a_3\equiv u_\alpha \sigma^{\alpha \beta} \sigma_{\beta \gamma}
u^\gamma\, , \label{a3}
\end{equation}
may be rewritten as
\begin{equation}
a_3 = -\ U_\alpha U^{\alpha}  \label{a3_2}\, ,
\end{equation}
with
\begin{equation}
U^{\alpha}\equiv \sigma^{\alpha \beta} u_{\beta}\, . \label{U}
\end{equation}

Consider the momentum vector $P^{\alpha}$ which is always timelike
(due to mass conservation). We now show that $U_{\alpha}$ is a
spacelike vector because it is orthogonal to $P^{\alpha}$. The
orthogonality relation reads
\begin{equation}
P^{\alpha} U_{\alpha}= P^{\alpha} {\sigma_{\alpha}}^{ \beta}
u_{\beta} =  \frac{D P^\beta}{D\lambda}\ u_{\beta}= 0\, ,
\label{PU2}
\end{equation}
where we have used \eqref{tetradev} for $e_{{(0)}}^{\mu}= P^\mu/{m}$, and the fact that $u_{\beta}$ is orthogonal to $D
P^\beta/{D\lambda}$ as it can be easily seen from \eqref{momentummotion}. Therefore, $U_\alpha U^\alpha < 0$, and then $a_3$ is positive.

It is enough to choose $AB < 0$ to end the proof that the Lagrangian
\eqref{lagsp0} is well defined irrespective of the sign of $a_1$.
Even if $AB \geq 0$ there are regions where Lagrangian
\eqref{lagsp0} is well defined for $a_1 < 0$, but for $AB < 0$, it
is well defined everywhere.

\section{Exact solution}

The equatorial motion of a top in a Schwarzschild field background
may be solved exactly. The Schwarzschild line element is $
ds^2=g_{tt}  dt^2+g_{rr}dr^2+g_{\theta\theta}d\theta^2+g_{\phi\phi}
d\phi^2$, where $g_{tt}=c^2(1-2r_0/r)$,
$g_{rr}=\left(1-2r_0/r\right)^{-1}$, $g_{\theta\theta}=-r^2$,
$g_{\phi\phi}=-r^2\sin^2\theta$, and $2 r_0$ is the Schwarzschild
radius. From now on we reinsert explicitly $c$ in all the
expressions. The general equations \eqref{momentummotion},
\eqref{spinmotion} were written in Ref.~\cite{hojman1} along with
\eqref{constraint}, \eqref{mass}, \eqref{spin}, \eqref{cxi} and
\eqref{gauge2} for the four Killing vectors of the Schwarzschild
metric. In this article, we restrict ourselves to the motion in the
plane defined by $\cos\theta=0$. If the top is initially in that
plane and $\dot\theta=0$, then it remains in the equatorial plane
\cite{hojman1}, in which $\theta=\pi/2$ and $P^\theta=0$. This
reduction is possible due to the fact that the direction of angular
momentum is conserved, so two of the four Killing vector
conservation laws are used to restrict the motion to the equatorial
plane.

It is convenient to define the dimensionless parameter
\begin{equation}
 \eta=\frac{J^2 r_0}{m^2c^2 r^3}\, ,
\label{eta}
\end{equation}
where $J=\hbar/2$ is the top's spin (as well as the neutrino's
spin). Thus, the set of equations
\eqref{momentummotion}--\eqref{gauge2} (including the two remaining
Killing vector conservation laws, energy $E$ and total angular
momentum magnitude $j$ in addition to the conservation of mass and
spin) may be solved exactly to yield \cite{hojman1}
\begin{equation}
 P_\phi=\frac{-j\pm E J/(m c^2)}{1-\eta}\, ,
\label{pfi}\end{equation}
\begin{equation}
 P_t=\frac{E\mp j J r_0/(m r^3)}{1-\eta}\, ,
\label{pt}\end{equation}
and, from $P_\mu P^\mu=m^2c^2$, we get
\begin{equation}
 P^r=\pm \left[\frac{P_t^2}{c^2}-\left(\frac{P_\phi^2}{r^2}+m^2c^2\right)\left(1-\frac{2r_0}{r}\right)\right]^{1/2}\, .
\label{pr1}\end{equation}

 We can now solve for the velocities. To this end, we
use two of the equations of motion \eqref{spinmotion}. In the plane
defined by $\theta=\pi/2$, these equations become
\begin{eqnarray}\label{}
  \frac{D S^{tr}}{D\lambda}&=&P^t \dot r-P^r\nonumber\\
  &=&\frac{S^{\phi r}P_\phi}{P_t^2}\frac{D P_t}{D\lambda}-\frac{D S^{\phi r}}{D\lambda}\frac{P_\phi}{P_t}-\frac{S^{\phi r}}{P_t}\frac{D P_\phi}{D\lambda}\, ,
\end{eqnarray}
and
\begin{eqnarray}\label{}
  \frac{D S^{t\phi}}{D\lambda}&=&P^t \dot \phi-P^\phi\nonumber\\
  &=&-\frac{S^{\phi r}P_r}{P_t^2}\frac{D P_t}{D\lambda}+\frac{D S^{\phi r}}{D\lambda}\frac{P_r}{P_t}+\frac{S^{\phi r}}{P_t}\frac{D P_r}{D\lambda}\, ,
\end{eqnarray}
which may be solved for $\dot r$ and $\dot\phi$. To perform this
task, we use the equations of motion \eqref{momentummotion} and
\eqref{spinmotion}, the relations between the spin and momentum
\cite{hojman1}
\begin{equation}
S^{tr}=-\frac{S^{\phi r}P_\phi}{P_t}\, ,
\end{equation}
\begin{equation}
S^{t\phi}=\frac{S^{\phi r}P_r}{P_t}\, ,
\end{equation}
which are consequences of the constraint \eqref{constraint}, and the
condition
\begin{equation}
\left(S^{\phi r}\right)^2=\frac{J^2 \left(P_t\right)^2}{m^2 r^2}\, ,
\end{equation}
which is derived from Eqs.~\eqref{constraint}, \eqref{mass} and \eqref{spin}.

Taking these results into account, the velocities turn out to be
\begin{equation}
 \dot\phi=\frac{c^2}{r^2}\left(1-\frac{2 r_0}{r}\right)\left(\frac{2\eta+1}{\eta-1}\right)\left(\frac{P_\phi}{P_t}\right)\, ,
\label{phipunto}\end{equation}
\begin{equation}
 \dot r=c^2\left(1-\frac{2 r_0}{r}\right)\left(\frac{P^r}{P_t}\right)\, .
\label{rpuntop1}\end{equation}
Finally, we get
\begin{equation}
\frac{d\phi}{dr}=\left(\frac{2\eta+1}{\eta-1}\right)\left(\frac{P_\phi}{r^2
P^r}\right)\, .
\label{dphidr}
\end{equation}

It is worth noting that the three preceding expressions coincide
with the usual results for geodesic motion when $J^2=0$ (and
therefore $\eta=0$). Once the solutions are spelled out, we can find
one of the main results of this work. From expressions
\eqref{phipunto} and \eqref{rpuntop1}, we find
\begin{equation}
 \left(\frac{ds}{cdt}\right)^2= \frac{g_{tt}}{c^2}+g_{rr}\left(\frac{\dot r}{c}\right)^2+g_{\phi\phi}\left(\frac{\dot \phi}{c}\right)^2=\frac{m^2}{(P^t)^2}(1-\Lambda)\, ,
\label{dsdt}
\end{equation}
where $c^2P^t=(1-2r_0/r)^{-1}P_t$, and we define the superluminal parameter $\Lambda$ as
\begin{equation}
 \Lambda=\frac{3 \eta (2+\eta)}{m^2 c^2 r^2 (1-\eta)^2}(P_\phi)^2>0\, .
 \label{lambdasuperluminal}
\end{equation}
From \eqref{dsdt} is straightforward to realize that it is possible
that $ds^2<0$ for some of the solutions, at least in part of the
top's trajectories. It is clear that the contribution of the
$\Lambda$ parameter is important for small mass particles, such as
neutrinos, because the $\Lambda$ dependence on the particle mass
behaves as $m^{-4}$.

In the next section we study the trajectory of small mass particles
under the approximation in which the top moves
far away from the Schwarzschild horizon, i.e., for $r \gg r_0$. We find the
conditions that the mass, energy and total angular momentum of the
top must satisfy in order to produce superluminal motion in a
segment of its trajectory.

\section{Superluminal motion for $r  \gg r_0$}

The main purpose of this section is to show that there are some
particle trajectories such that $\Lambda<1$ in part of the path and
$\Lambda>1$ in the rest of it. Knowing that $J=\hbar/2$ is the spin
of the top, superluminal behavior imposes a condition on the
mass $m$, the total energy $E$, and the total angular momentum $j$.

We look for a solution such that the particle is moving at distances
$r$ much larger than the Schwarzschild radius of the black hole,
$r_0/r\ll 1$, i.e., we only consider the motion of the particle in
a weak gravitational field. Besides, to focus in regimes in which
$\Lambda$ is relevant, we restrict ourselves to the case of a small
mass particle, such that its total energy is larger than its rest
mass energy, $E \gg mc^2$. Both assumptions imply that $\eta\ll 1$.

In the weak field and small mass approximations, taking the upper
signs in the preceding solutions \eqref{pfi} and \eqref{pt}, the
momenta become $P_\phi=\left(1+\eta\right)\left(-j+ EJ/m
c^2\right)$, $P_t=\left(1+\eta\right)\left(E- j J r_0/m r^3\right)$,
and
\begin{equation}
 P^r=\pm\left[\frac{E^2}{c^2}-\frac{1}{r^2}\left(j-\frac{EJ}{mc^2}\right)^2\right]^{1/2}\, .
\label{pr2}\end{equation}

In order to study the particle motion, we assume that initially the
particle approaches the central body from infinity and it remains
always at a distance $r\gg r_0$. Thus, the orbit of the particle in
the equatorial plane can be characterized by two values of $r$ in
the trajectory, $r_c$ and $r_R$. The first one is the value of the
orbit's radius in which the particle's velocity reaches the speed of
light, i.e., the point in which $\Lambda=1$. The second one is the
return point of the particle's orbit, and it is defined as the point
when its radial momentum vanishes, $P^{r}=0$.

By imposing the condition
$\Lambda=1$, it is straightforward to obtain that the value of $r_c$ is given by
\begin{equation}
 r_c=\left[\frac{6J^2r_0}{m^4c^4}\left(j-\frac{EJ}{mc^2}\right)^2\right]^{1/5}\, .
\label{rc}\end{equation}

On the other hand, we can calculate the return point for the
particle's trajectory $r_{R}$ solving $P^{r}=0$ from \eqref{pr2}. We get
\begin{equation}
 r_R= \frac{cj}{E}\left(1 - \frac{JE}{j\, mc^2}\right)\, .
 \label{returnpoint}
\end{equation}

The particle reaches the speed of light in $r_c$ (because
$\Lambda(r_c)=1$ and therefore $ds^2=0$). However, to have
superluminal motion after that point requires that $\Lambda(r_R)>1$
(implying $ds^2<0$). Using Eq.~\eqref{returnpoint} in
\eqref{lambdasuperluminal}, this implies a condition that the mass,
the energy and the total angular momentum of the particle must
fulfill
\begin{equation}
  6J^2E^5 r_0>m^4 c^9\left(j-\frac{EJ}{mc^2}\right)^3\, .
  \label{condicionsuperlu}
\end{equation}

The above expression shows the relation that $m$, $E$ and $j$ satisfy to achieve superluminal motion. Notice that the condition
cannot be met by spinless particles. Knowing the energy of the
particle, it is possible to use \eqref{condicionsuperlu} to estimate
the mass and total angular momentum of the particle if superluminal
motion is ever detected.

Interestingly, one can prove that condition \eqref{condicionsuperlu}
is equivalent to $r_R < r_c$. This means that when the particle has
a "ballistic" trajectory around the black hole, it can always be
superluminal in some part of its orbit, nearest to $2r_0$, for
appropriate $m$, $E$ and $j$. Clearly its speed will not be constant
along the trajectory, because in the first part of the trip, the
gravitational field speeds it up until it reaches $r_R$, slowing it
down afterwards.

\section{Conclusions}

We present an exact solution to the non geodesic equations of motion
in a Schwarzschild gravitational background that allows for
superluminal propagation of massive spinning test particles. The
superluminal motion depends strongly of the inverse of the mass of
the particle. This is, of course, relevant for small mass particles
such as neutrinos. We have shown that a consistent general
relativistic theory which allows for superluminal propagation is
possible.

Furthermore, the superluminal propagation effect presented here can be achieved in weak gravitational fields,
as for example, in the surface of the Earth. In the presence of stronger gravitational fields
(which can be easily found in astrophysical context) this effect will be much enhanced.

It is worthwhile mentioning that the aforementioned effect depends
strongly on the fact that we deal with small mass particles, as the
relevant $\Lambda$ parameter has an $m^{-4}$ dependence, and it is
therefore very unlikely that superluminality could be detected for
particles other than neutrinos.

Finally, we would like to mention that the equation of motion
\eqref{momentummotion} can be rigorously generalized to include the
gravitational self-force of the tops \cite{wald}, showing that these
corrections also modify the geodesic paths. However, the new forces
are proportional to the particle mass and therefore their effects
for small mass particles are negligible compared with those
presented in this work.

\appendix
\section{}

The mass conservation law may be obtained as follows.
Rewrite the velocity vector $ u^\nu$ as
\begin{eqnarray}
 u^\nu&=&\frac{1}{m^2}\left(P_\mu P^\mu\right)u^\nu+\frac{1}{m^2}\left(P_\mu u^\mu-P_\mu u^\mu\right)P^\nu\, \nonumber\\
&=&\frac{1}{m^2}\left(P^\mu u^\nu-P^\nu
u^\mu\right)P_\mu+\frac{1}{m^2}P_\mu u^\mu P^\nu\, .
\end{eqnarray}

We use the equation of motion for the spin tensor \eqref{spinmotion}
to deal with the first term on the right hand side of the previous
equation. Then
\begin{eqnarray}
 m^2u^\nu&=&P_\mu\frac{D S^{\mu\nu}}{D\lambda}+P_\mu u^\mu P^\nu\, ,\nonumber\\
&=&-\frac{D P_\mu}{D\lambda}S^{\mu\nu}+P_\mu u^\mu P^\nu\, ,
\end{eqnarray}
where we have used the constraint $S^{\mu\nu}P_\mu=0$. We multiply
by $D_\lambda P_\nu$, and due to the fact that $D_\lambda P_\mu
S^{\mu\nu} {D_\lambda}P_\nu\equiv 0$, we find
\begin{equation}
m^2 u^\nu\frac{D P_\nu}{D\lambda}=P_\mu u^\mu P^\nu \frac{D
P_\nu}{D\lambda}\, .
\end{equation}

However, using the equation of motion for $P_\nu$, we find that
$u^\nu D_\lambda P_\nu=0$, and then we get the condition
\begin{equation}
 P^\nu \frac{D P_\nu}{D\lambda}=0\, ,
\end{equation}
which implies that $ m^2$ is constant.

\section{}
In this Appendix we show how to implement the choices
\begin{equation}
{e_{(0)}}^{\mu}=P^\mu/m\, . \label{Appgauge}
\end{equation}
using a different approach to the one presented in \cite{hr}.

Let's start by defining $\bar{S}_{({\alpha \beta})}$ by
\begin{equation}
\bar{S}_{({\alpha \beta})} \equiv {e_{(\alpha) \mu}}S^{\mu
\nu}{e_{(\beta)\nu}}\, . \label{sbar}
\end{equation}
It is a straightforward matter to realize that the six quantities
$\bar{S}_{({\alpha \beta})}$ are constants of motion
\begin{equation}
\frac{D \bar{S}_{({\alpha \beta})}}{D\lambda} =0\, ,
\label{sbarconst}
\end{equation}
because of \eqref{spinmotion} and \eqref{tetradev}. We can now
choose
\begin{equation}
\bar{S}_{(0 i)} =0\, ,\label{sbar0i}
\end{equation}
as three initial conditions which are, of course, preserved in time
because of \eqref{sbarconst}. Hanson and Regge call these conditions ``invariant relations''.

Conditions \eqref{sbar0i} imply that the vector ${e_{(0) \mu}}$ is a
null eigenvector of the spin matrix $S^{\mu \nu}$, i.e.,
\begin{equation}
S^{\mu \nu}{e_{(0) \nu}} =0\, . \label{nullev}
\end{equation}
The spin matrix $S^{\mu \nu}$ has even rank (because it is
antisymmetric). Due to the fact that $S^{\mu \nu}P_\nu=0$, its rank
is not four, so it must be two (otherwise it would be zero,
rendering it trivial). If its rank is two, it must have two null
eigenvectors, which are the momentum vector $P^\mu$ and the Pauli
Lubanski vector $W^\mu \equiv \frac{1}{2}\epsilon^{\mu \nu \alpha
\beta} S_{\alpha \beta} P_\nu$, as one can easily prove (see also
\cite{hr}).

Therefore, the vector ${{e_{(0)}}^{\mu}}$ may be expressed as a
linear combination of the momentum vector $P^\mu$ and the Pauli
Lubanski vector $W^\mu$
\begin{equation}
{{e_{(0)}}^{\mu}}= \rho \frac{P^\mu}{m} + \tau W^\mu \,  .
\label{e0pw}
\end{equation}

Hanson and Regge \cite{hr} construct the Hamiltonian theory of the
top (pages 523 and following of reference \cite{hr}) using Dirac's
method \cite{dirac}. There they handle the Tulczyjew constraint
$S^{\mu \nu}P_\nu=0$ by extracting its first class content  $\Phi_2$
(in Dirac's terminology) with
\begin{equation}
\Phi_2 \equiv \frac{1}{2}\epsilon^{\mu \nu \alpha \beta} S_{\mu \nu
}S_{\alpha \beta}\approx 0\, , \label{constraint2}
\end{equation}
where the sign ``$\approx 0$'' is read ``weakly equal to zero''.
This means that the constraint $\Phi_2$ vanishes, but its Poisson
bracket relations with some dynamical variables are different from
zero. Loosely speaking, Dirac's method provides one way to mend this
contradiction by redefining the Poisson brackets relations. The new
brackets are called Dirac brackets in his honor. The Dirac brackets
of the constraints (with any dynamical variable) are
identically zero.

Due to the fact that $\Phi_2$ is first class, it generates gauge
transformations \cite{dirac}. Therefore a gauge (associated to it)
may be chosen. If the Poisson bracket of a dynamical variable $A$
with $\Phi_2$ is such that $[A,\Phi_2]\approx 0$, then $A$ is
invariant under $\Phi_2$, i.e., is gauge invariant. So, in order to
choose a gauge associated to $\Phi_2$ one needs to find a variable
$B$ such that its Poisson bracket with $\Phi_2$ be different from zero.

Use \eqref{e0pw} to get $\rho$
\begin{equation}
\rho= {{e_{(0)}}^{\mu}} \frac{P_\mu}{m}\, , \label{rho}
\end{equation}
and consider the Poisson bracket relations, equations (3.11) of
reference \cite{hr}
\begin{equation}
\left[{{e_{(\gamma)}}^{\mu}},S^{\alpha \beta}\right]= {{e_{(\gamma)}}^{\alpha}}
g^{\mu \beta} - {{e_{(\gamma)}}^{\beta}}g^{\mu \alpha}\, ,
\label{pbes}
\end{equation}
or
\begin{equation}
\left[{{e_{(\gamma)}}^{\mu}},S_{\alpha \beta}\right]= {{e_{(\gamma)
{\alpha}}}} {\delta^{\mu}}_{\beta} - {{e_{(\gamma) {\beta}}}}
{\delta^{\mu}}_{\alpha}\, , \label{pbesdown}
\end{equation}
(it is perhaps worth mentioning that there is a change in notation,
the role of Hanson and Regge's  ${\Lambda_{\mu}}^{\nu}$ matrix
\cite{hr} is played by the tetrad vectors \cite{hojman1,hojman2}
${{e_{(\mu)}}^{\nu}}$ here). Now, it is straightforward to realize
that $[\rho,\Phi_2]=4 {e_{(0){\mu}}} W^\mu\neq 0$ (if
${e_{(0){\mu}}} W^\mu=0$, the proof ends here). 

Squaring \eqref{e0pw} one gets
\begin{equation}
1 = {\rho}^2 -{\tau}^2 m^2 J^2\, . \label{e02}
\end{equation}

We may, therefore, choose the gauge
\begin{equation}
\rho = 1\, , \label{rhogauge}
\end{equation}
which means that $\tau =0$, thus ending the proof.

The same result may be achieved by computing $[\tau,\Phi_2]$ and
choosing the gauge $\tau=0$.

\begin{acknowledgments}

F.A.A. thanks the CONICyT-Chile for a BecasChile Postdoctoral Fellowship.

\end{acknowledgments}

\end{document}